\documentclass{article}

\usepackage{PRIMEarxiv}

\usepackage[utf8]{inputenc} 
\usepackage[T1]{fontenc}    
\usepackage{hyperref}       
\usepackage{url}            
\usepackage{booktabs}       
\usepackage{amsfonts}       
\usepackage{nicefrac}       
\usepackage{microtype}      
\usepackage{lipsum}
\usepackage{fancyhdr}       
\usepackage{graphicx}       
\usepackage{booktabs}
\usepackage{multirow}
\usepackage{fancyhdr}
\usepackage{xcolor}
\usepackage{float} 
\title{Gender Representation in Brazilian \\ Computer Science Conferences
\thanks{\textit{\underline{Citation}}: 
\textbf{DAL PIZZOL, Natália; BARBOSA, Eduardo Dos Santos; MUSSE, Soraia Raupp. Gender Representation in Brazilian Computer Science Conferences. In: WOMEN IN INFORMATION TECHNOLOGY (WIT), 16. , 2022, Niterói. Anais [...]. Porto Alegre: Sociedade Brasileira de Computação, 2022 . p. 67-76. ISSN 2763-8626. DOI: https://doi.org/10.5753/wit.2022.222939}} 
}

\author{
  Natália Dal Pizzol, Eduardo Dos Santos Barbosa and Soraia Raupp Musse \\
  School of Technology \\
  Pontifical Catholic University of Rio Grande do Sul \\
  Porto Alegre, Brazil\\
  \texttt{{natalia.pizzol,eduardo.santos94}@edu.pucrs.br, soraia.musse@pucrs.br}
}

\begin{document}
\maketitle

\begin{abstract}
This study presents an automated bibliometric analysis of 6569 research papers published in thirteen Brazilian Computer Science Society (SBC) conferences from 1999 to 2021. Our primary goal was to gather data to understand the gender representation in publications in the field of Computer Science. We applied a systematic assignment of gender to 23.573 listed papers authorships, finding that the gender gap for women is significant, with female authors being under-represented in all years of the study.\end{abstract}


\noindent \color{red}WORK DRAFT ONLY. Please cite final version. Final version published as \color{black}: DAL PIZZOL, Natália; BARBOSA, Eduardo Dos Santos; MUSSE, Soraia Raupp. Gender Representation in Brazilian Computer Science Conferences. In: WOMEN IN INFORMATION TECHNOLOGY (WIT), 16. , 2022, Niterói. Anais [...]. Porto Alegre: Sociedade Brasileira de Computação, 2022 . p. 67-76. ISSN 2763-8626. DOI: https://doi.org/10.5753/wit.2022.222939.

\section{Introduction}

Although female figures have played a crucial role in the development and advancement of Computer Science (CS), the field has become predominantly male~\cite{Wang2021GenderTI}. This gender disparity is considered a global phenomenon, and one way that it presents itself in Brazil is in the low rates of enrollment for female students in Computer Science degrees. The 2019 Statistics on College Education Report, published by the  Brazilian Computer Society (SBC)\footnote{\url{https://www.sbc.org.br/documentos-da-sbc/category/133-estatisticas}}, shows that women represent only 11.51\% of incoming CS students. 

Despite increased discussions and concerns about the inclusion of women in Computer Science communities, evidence has not yet been obtained that can clearly state how this disparity in the gender proportion impacts research publications in Brazilian conferences. Research publications are the main instrument to disseminate scientific knowledge~\cite{Holman2018TheGG}, and in the case of Brazilian CS researchers, works are often published in conference proceedings~\cite{albertini}. 

According to the Brazilian Computer Society's list of events\footnote{\url{https://www.sbc.org.br/eventos/eventos-realizados}}, many conferences in Computer Science have been created in the last 20-30 years in Brazil. The exact number of conferences active in 2022 is particularly difficult to estimate if we consider regional or local workshops. However, if we select conferences supported by the Brazilian Computer Society, for which many papers are available in public data sets, we can achieve the objective of finding Brazilian conferences to evaluate the diversity in gender representation of published authors. 

The goal of this work is to present an analysis of the diversity and representation of gender in Brazilian conferences over time by conducting a bibliometric study based on systematic assignment of gender. This article is organized as follows: Section~\ref{sec:relatedwork} presents works related to the topic of diversity and female representation in Computer Science. Section~\ref{sec:methodology} outlines the proposed methodology for the study, including data collection, gender inference and implementation. Section~\ref{sec:results} details the results obtained in this research. Finally, Section~\ref{sec:finalconsiderations} presents our final considerations.

\section{Related Work} \label{sec:relatedwork}

The lack of diversity and absence of a variety of different voices in Science, Technology, Engineering and Mathematics (STEM), has been a frequent topic of study in recent years. Haines et al.~\cite{hainesETAL2020} have investigated the influence of the researcher's gender on their research topic by analyzing birdsong literature. In animal behavior studies, birdsong was studied as a primarily male trait, however, they have found that women working on the field are more likely to be the authors of female birdsong articles. The significant contributions that women have made to the field of female birdsong studies suggest that diversity in academia can foster new scientific ideas, maximizing the value and quality of research.

Nevertheless, even with studies showing the importance of diversity in science, gender disparities persist. In 2013, Larivière et al.~\cite{Cassidy2013} conducted a global multidisciplinary study on scientific publications indexed in Thomson Reuters Web of Science databases between 2008 and 2012. They found that, globally, female authorship accounted for fewer than 30\% of publications, and that articles with women in dominant author positions tend to receive fewer citations. Their results showed how prevalent gender inequality is in STEM.

In the field of Computer Science, several studies focused on mapping global gender representation. Wang et al.~\cite{Wang2021GenderTI} analyzed gender trends in Computer Science literature with a sample of 11.8M articles published from 1970 to 2019 that were indexed in the Semantic Scholar literature corpus. Their findings show that, although the proportion of female authors is increasing, there is still a notable gender gap in the academic authorship of CS research.

The overall under-representation of women in CS research is well-established, but other studies have been conducted to scan the distribution of this inequality across CS subfields. Cheon et al.~\cite{Cheong2021} have analyzed nine subfields, and found that female authors are outnumbered in each one of them, and that the fields of Artificial Intelligence, Information Security, Computer Vision, Machine Learning, and Systems Architecture represent the less gender-diverse areas, with an approximate 5:1 ratio of male to female authors.

Ribeiro et al.~\cite{sbcassociados} have conducted a quantitative analysis of SBC members, investigating their gender, location, type of membership, and areas of interest in CS. They found that 77,71\% of SBC members are male and 21,67\% are female. Regarding areas of interest in CS, they found that there is a bigger proportion of women interested in the areas of Information Systems, Multimedia and Hypermedia Systems, Collaborative Systems, Informatics in Education, and Human Computer Interaction, while the least favored areas by women are Computational Architecture and High Performance Processing, Distributed Systems, Integrated Circuit Design, Computer Networks and Distributed Systems, and Algorithms, Combinatorics and Optimization.

Arruda et al.~\cite{arruda} have analyzed 886 publications from 2000 to 2006 authored by Brazilian researchers. Their work classifies researchers into CS subfields, and suggests that female scientists tend to concentrate in the areas of Artificial Intelligence, Collaborative Systems, Computer in Education, and Human-Computer Interfaces. This research, however, shows some limitations since they were working with a small sample of authors.

Although several contributions have been made to the study of gender representation in CS, fewer articles focus on analyzing Brazil specifically. In order to contribute to this investigation, the main research question in our work concerns gender representation in Brazilian Computer Science conferences. Next section presents the applied methodology.

\section{Methodology} \label{sec:methodology}

The primary goal of this study is to assess the gender diversity and representation of authors who contribute to advancing computer science research in Brazil. To that end, we analyzed publications from thirteen SBC conferences, as listed in Table~\ref{tabela_confs}. Those 13 conferences were selected from the total of 31 events listed in the SBC Open Library of publications\footnote{\url{https://sol.sbc.org.br/index.php/anais/confs}}. There were three main criteria for the selection of the conferences in this work: 
\begin{itemize}
\item the conference must be sponsored by the Brazilian Computer Society\footnote{\url{sbc.org.br}}. This criteria aims to avoid local or regional and small workshops.
\item the conference should have had at least 20 editions. This criteria was used in order to select more consolidated conferences in Brazil.
\item the conference should be indexed in either the Scopus or DBLP databases. Finally, this criteria was chosen to identify papers that have more scientific visibility.
\end{itemize}

More details on the data collection and gender inference processes are provided in the following subsections.

\subsection{Data Collection and Processing} 
\label{datacollection}
There are several available databases to consult computer science publications, some of the most commonly used are DBLP, Scopus, and Web of Science. This study was conducted using DBLP\footnote{\url{https://dblp.org/}}, due to its high index of unique articles~\cite{cavacini}, programmer-friendly API, and free service. 
We used DBLP API's query for venues, which returns a list of every indexed publication for a determined conference, journal, etc. We applied the names of the selected conferences, presented in Table~\ref{tabela_confs}, as query parameters. This DBLP API query returned the associated metadata for each publication, allowing for the retrieval of title, author and coauthors, year of publication, type of publication (such as conference or workshop papers), DOI (Digital Object Identifier) and venue information.

The next step was to sort and filter the publication information. We chose to exclude publications that did not contain DOI information in the associated metadata, as the DOI is one of the most reliable identifiers of a publication. The publication's DOI ensured that no articles in our list were repeated, and increased the transparency of the research since all works could be referenced and located. 

\begin{table} [ht]
\centering
\small
\caption{SBC conferences analyzed in the study.}
\label{tabela_confs}
\begin{tabular}{llc} 
\toprule
\textbf{Conference Name} & \textbf{Reference} & \multicolumn{1}{l}{\textbf{Editions*}} \\ 
\hline
Brazilian Symposium on Human Factors in Computational Systems & IHC & 20 \\
Symposium on Computer Architecture and High Performance Computing & SBAC-PAD & 33 \\
Brazilian Symposium on Databases & SBBD & 36 \\
Brazilian Symposium on Integrated Circuits and Systems Design & SBCCI & 33 \\
Brazilian Symposium on Software Engineering & SBES & 35 \\
Brazilian Symposium on Computer Games and Digital Entertainment & SBGAMES & 20 \\
Brazilian Symposium on Programming Languages & SBLP & 25 \\
Brazilian Symposium on Formal Methods & SBMF & 24 \\
Brazilian Symposium on Software Quality & SBQS & 20 \\
Brazilian Symposium on Computer Networks and Distributed Systems & SBRC & 39 \\
SIBGRAPI Conference on Graphics, Patterns and Images & SIBGRAPI & 33 \\
Symposium on Virtual and Augmented Reality & SVR & 23 \\
Brazilian Symposium on Multimedia and the Web & WebMedia & 22 \\
\bottomrule
\multicolumn{2}{l}{*As of March 2022.} & 
\end{tabular}
\end{table}

Next, we filtered the publications to exclude those with the type listed as ``Editorship", to ensure that non-scientific works would not be included in the final analysis. After the filtering process, the data frame contained 6569 publications with 23573 authorships. Table~\ref{autoresdat} presents information regarding the number of articles and authorships for each of the analyzed conferences.

\begin{table} [ht]
\footnotesize
\centering
\caption{Data collected per conference.}
\label{autoresdat}
\begin{tabular}{lrrrrl} 
\toprule
\multicolumn{1}{c}{\textbf{Conference}} & \multicolumn{1}{c}{\begin{tabular}[c]{@{}c@{}}\textbf{Total }\\\textbf{ Publications}\end{tabular}} & \multicolumn{2}{c}{\begin{tabular}[c]{@{}c@{}}\textbf{Filtered }\\\textbf{Publications}\end{tabular}} & \multicolumn{1}{c}{\begin{tabular}[c]{@{}c@{}}\textbf{Number of}\\\textbf{ Authorships}\end{tabular}} & \textbf{Years Indexed on DBLP} \\ 
\cline{3-4}
\multicolumn{1}{c}{} & \multicolumn{1}{c}{} & \multicolumn{1}{c}{Count} & \multicolumn{1}{c}{\%} & \multicolumn{1}{c}{} &  \\ 
\hline
IHC & 620 & 518 & 83.5 & 1785 & 2006, 2008, 2010-2021 \\
SBAC-PAD & 711 & 711 & 100 & 2760 & 2002-2021 \\
SBBD & 680 & 109 & 16 & 420 & 1999-2021 \\
SBCCI & 828 & 827 & 99.8 & 2944 & 2003-2020 \\
SBES & 489 & 489 & 100 & 1940 & 2009-2021 \\
SBGAMES & 241 & 241 & 100 & 881 & 2009-2011, 2014-2015, 2017-2021 \\
SBLP & 112 & 112 & 100 & 352 & 2012-2021 \\
SBMF & 178 & 178 & 100 & 500 & 2009-2018, 2020-2021 \\
SBQS & 155 & 155 & 100 & 604 & 2018-2021 \\
SBRC & 603 & 525 & 87 & 1930 & 2014, 2015, 2017-2021 \\
SIBGRAPI & 1231 & 1231 & 100 & 4083 & 1999-2021 \\
SVR & 620 & 603 & 97.2 & 2322 & 2012-2021 \\
WEBMEDIA & 890 & 870 & 97.7 & 3052 & 2005, 2006, 2008, 2009, 2012-2021 \\
\bottomrule
\end{tabular}
\end{table}

\subsection{Gender Inference}
Considering that gender information is not available in most databases, including DBLP, one of the most reliable ways to infer an author's gender is by analyzing their name. We combined three different ways to assign gender to first names, the Gender API\footnote{\url{https://gender-api.com/}}, the Python package gender-guesser\footnote{\url{https://pypi.org/project/gender-guesser/}}, and the gender classification data made available by the Brazil.IO project\footnote{\url{https://brasil.io/dataset/genero-nomes/nomes/}}.
Gender API is an online service with an extensive database. Gender-guesser is an offline package with a more limited amount of names in its dictionary, however, its data was manually checked by native speakers of different countries, and therefore is presumed to be of high quality, as supported by Santamaría and Mihaljevic~\cite{Santamar_a_2018}. 

Brasil.IO's gender classification data reflects the self-informed gender of Brazilian residents as collected in the 2010 Brazilian Census\footnote{\url{https://censo2010.ibge.gov.br/nomes/\#/search}}, and is made available in a csv file containing the name's classification (male or female), its frequency of appearance as female and male, and the ratio (ranging from 0 to 1), which represents the confidence for the classification. We filtered the data to only include classifications for names that had a ratio of at least 0.9.

We extracted 23573 author names from the list of publications, split them into firs and last names, and used the first name string to first query gender-guesser. Gender-guesser assigns gender as unknown (for a name not found in the database), andy (androgynous names, i.e. names that have a similar probability to be male than to be female), male, female, mostly\_male, or mostly\_female. Then, for the names that were assigned gender as unknown by gender-guesser, we applied Gender API. Gender API returns a gender assignment with the possible values of male, female or unknown. Lastly,  we cross-referenced the names that were still classified as unknown with Brasil.IO's data, applying their classification for names that had a ratio of at least 0.9, in order to avoid ambiguity.

We were able to assign a gender for 91,88\% of names in the authorship list. It is important to note that research articles often have more than one author, and albeit coauthors might have different extents of contribution, for the purpose of this study, we considered all authors listed in the publication equally. In order to analyze gender representation in scientific productions, every author name listed was counted as one authorship, meaning that one author could have published more than once in the conferences and, for each instance, it would have counted as different authorships. The number of unique author entries and authorships is shown in Table~\ref{authorships}.

\begin{table} [ht]
\centering
\caption{Unique author entries and authorships by gender assignment.}
\label{authorships}
\begin{tabular}{lrr} 
\toprule
\textbf{Gender Assignment} & \textbf{Unique Author Entries} & \textbf{Number of Authorships} \\ 
\hline
Male & 8691 & 17140 \\
Female & 2000 & 4120 \\
Unknown & 1271 & 1914 \\
Mostly\_male & 189 & 382 \\
Mostly\_female & 13 & 17 \\
\bottomrule
\end{tabular}
\end{table}

By comparing the number of unique author entries with the authorships count, as detailed in Table~\ref{authorships}, we define the productivity factors $p_f$ and $p_m$. $p_f$, shown in Equation \ref{eq_pf}, denotes the productivity factor for female authors. $p_m$, show in Equation~\ref{eq_pm}, denotes the productivity factor for male authors. 

\begin{equation} \label{eq_pf}
    p_f =  \frac{Authorship(f+mf)}{Unique(f+mf)},
\end{equation}

where $f$ stands for gender assignment female and $mf$ stands for mostly\_female.

\begin{equation} \label{eq_pm}
    p_m= \frac{Authorship(m+mm)}{Unique(m+mm)},
\end{equation}

where $m$ stands for gender assignment male and $mm$ stands for mostly\_male.

\subsection{Implementation Details}

The main goal for the program's implementation was to make it as accessible and easy to replicate as possible. Our methodology was implemented using Python 3.10\footnote{\url{https://www.python.org/}} on a
11th Gen Intel(R) Core(TM) i5-1135G7 with a 16GB memory. We used the Python Requests library\footnote{\url{https://docs.python-requests.org/}} to retrieve conference data from the DBLP API. This information was stored in a csv file containing the associated metadata for each publication. The list was then filtered to remove publications that did not contain DOI information or whose type was listed as ``Editorship", as explained in Section~\ref{datacollection}. The data frame used for this analysis is available at GitHub repository\footnote{\url{https://github.com/Virtual-Humans-Lab/Gender-Representation-Analysis}}. 

We used the gender-guesser package version 0.4.0, and the Gender API service, following the ``Simple Usage"~Request, as detailed on the API's documentation \footnote{\url{https://gender-api.com/en/api-docs\#simple-usage}}.  
In order to reduce costs with the paid subscription of gender API, every name query and correspondent gender assignment were stored in a JSON file that the program would check before querying to the API, avoiding that multiple requests be made for names who appeared in the list repeatedly.

\section{Results}\label{sec:results}

This section presents the results obtained with our proposed methodology. All results are reflective of the DBLP repository as of March 19, 2022. We analyzed conferences ranging from 1999 to 2021, with a total of 6569 publications and 23573 authorships, as shown in Table~\ref{autoresdat}. Although gender inference based on name is not an infallible method, and raises ethical concerns for its exclusion of non-binary individuals, its application was relevant for this study, as we were able to gather gender data that would otherwise be inaccessible.

\begin{figure} [ht]
    \centering
    \includegraphics[width=17cm]{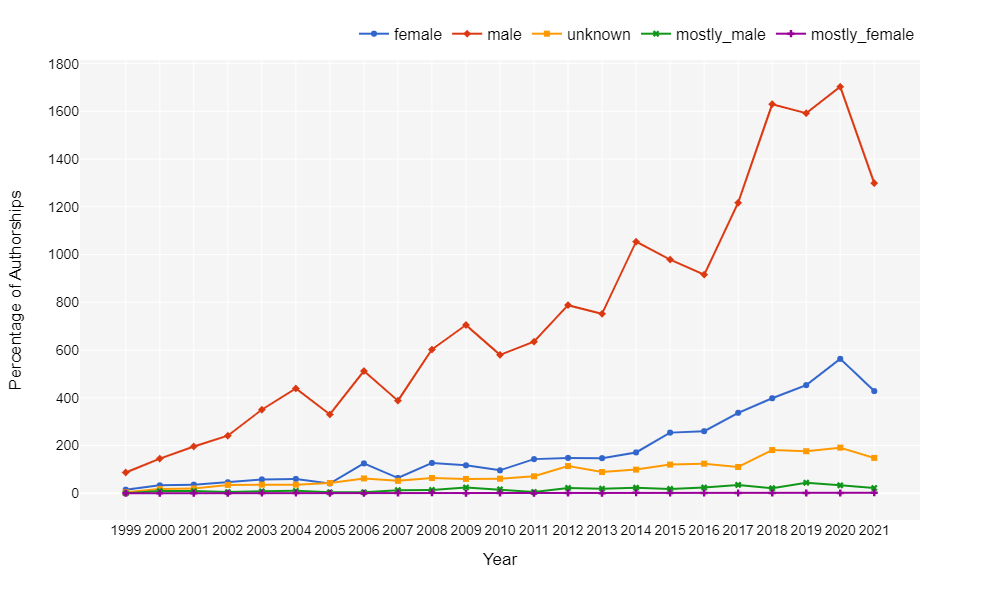}    
    \caption{Number of publications by gender per year in SBC's conferences.}
    \label{fig:gender_by_year}
\end{figure}

\begin{figure} [ht]
    \centering
    \includegraphics[width=17cm]{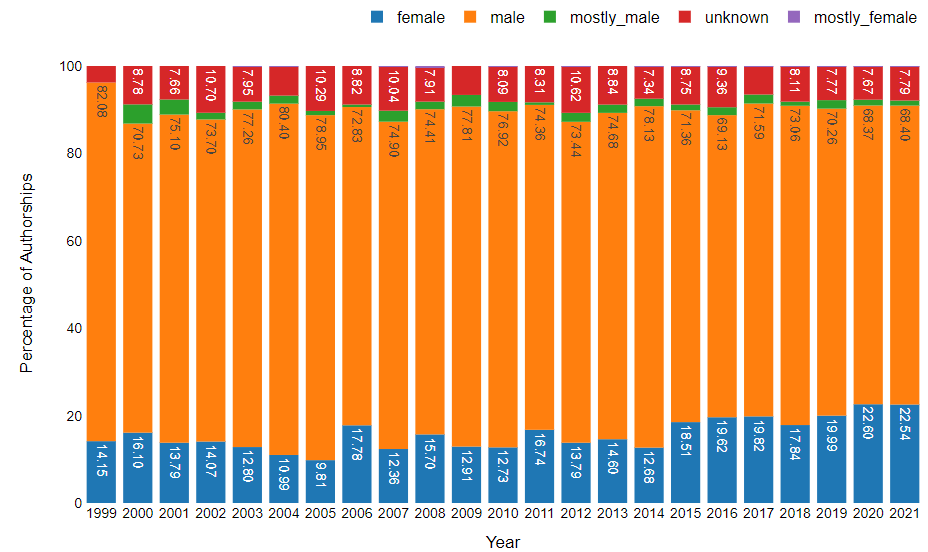}
    \caption{Gender representation by year in SBC's conferences.}
    \label{fig:gender_percent}
\end{figure}

\begin{figure} [ht]
    \centering
    \includegraphics[width=17cm]{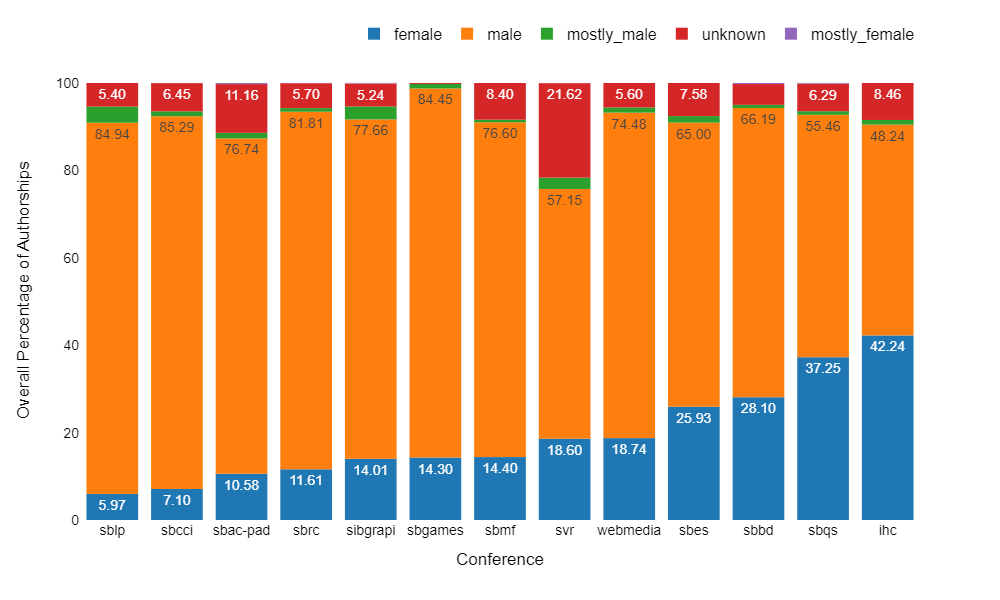}
    \caption{Overall gender representation by conference.}
    \label{fig: gender_confes}
\end{figure}

As the results of this work suggest, despite the number of female authors growing in the past 22 years, as of today, women are still underrepresented in Computer Science research published in SBC conferences, as shown in Figure \ref{fig:gender_by_year}. Furthermore, the gender gap for women in the analyzed conferences is significantly present in all years, as illustrated in Figure~\ref{fig:gender_percent}.
Concerning the overall gender representation by conference, as shown in Table~\ref{overallbyconf} and Figure \ref{fig: gender_confes}, we find that only two conferences, the Brazilian Symposium on Human Factors in Computational Systems (IHC) and the Brazilian Symposium on Software Quality (SBQS), have had at least 30\% female authorships in their publications. We also found that two conferences, the Brazilian Symposium on Programming Languages (SBLP) and the Brazilian Symposium on Integrated Circuits and Systems Design (SBCCI), have had under 10\% female representation.

\begin{table} [ht]
\centering
\small
\caption{Overall gender representation by conference.}
\label{overallbyconf}
\begin{tabular}{lrrrrrrrrrr} 
\toprule
\multicolumn{1}{c}{\textbf{Conference}} & \multicolumn{2}{c}{\textbf{Male}} & \multicolumn{2}{c}{\textbf{Female}} & \multicolumn{2}{c}{\textbf{Unknown}} & \multicolumn{2}{c}{\textbf{Mostly\_male}} & \multicolumn{2}{c}{\textbf{Mostly\_female}} \\ 
\cline{2-11}
 & \multicolumn{1}{c}{Count} & \multicolumn{1}{c}{\%} & \multicolumn{1}{c}{Count} & \multicolumn{1}{c}{\%} & \multicolumn{1}{c}{Count} & \multicolumn{1}{c}{\%} & \multicolumn{1}{c}{Count} & \multicolumn{1}{c}{\%} & \multicolumn{1}{c}{Count} & \multicolumn{1}{c}{\%} \\ 
\hline
IHC & 861 & 48.24 & 754 & 42.24 & 151 & 8.46 & 19 & 1.06 & 0 & 0 \\
SBAC-PAD & 2118 & 76.74 & 292 & 10.58 & 308 & 11.16 & 35 & 1.27 & 7 & 0.25 \\
SBBD & 278 & 66.19 & 118 & 28.10 & 20 & 4.76 & 3 & 0.71 & 1 & 0.24 \\
SBCCI & 2511 & 85.29 & 209 & 7.10 & 190 & 6.45 & 33 & 1.12 & 1 & 0.03 \\
SBES & 1261 & 65.00 & 503 & 25.93 & 147 & 7.58 & 29 & 1.49 & 0 & 0 \\
SBGAMES & 744 & 84.45 & 126 & 14.30 & 2 & 0.23 & 9 & 1.02 & 0 & 0 \\
SBLP & 299 & 84.94 & 21 & 5.97 & 19 & 5.40 & 13 & 3.69 & 0 & 0 \\
SBMF & 383 & 76.60 & 72 & 14.40 & 42 & 8.40 & 3 & 0.60 & 0 & 0 \\
SBQS & 335 & 55.46 & 225 & 37.25 & 38 & 6.29 & 5 & 0.83 & 1 & 0.17 \\
SBRC & 1579 & 81.81 & 224 & 11.61 & 110 & 5.70 & 16 & 0.83 & 1 & 0.05 \\
SIBGRAPI & 3171 & 77.66 & 572 & 14.01 & 214 & 5.24 & 121 & 2.96 & 5 & 0.12 \\
SVR & 1327 & 57.15 & 432 & 18.60 & 502 & 21.62 & 60 & 2.58 & 1 & 0.04 \\
WEBMEDIA & 2273 & 74.48 & 572 & 18.74 & 171 & 5.60 & 36 & 1.18 & 0 & 0 \\
\bottomrule
\end{tabular}
\end{table}

By analyzing the productivity factors $p_f$ and $p_m$, shown in Equations \ref{eq_pf} and \ref{eq_pm}, we see that each female author has published an average of 2.06 times in the analyzed conferences, while male authors published an average of 1.97 time. This suggests that the productivity of female authors is on par with that of male authors for the analyzed conferences. The overall gender classifications are show in \ref{authorships}. These numbers indicate that there are $4.16$ times more male authors than females, further showing that there is an urgent need to find strategies and policies to remedy this gender gap.


\section{Final Considerations} \label{sec:finalconsiderations}

We performed an analysis of the Computer Science literature output from Brazilian conferences to investigate gender representation. Our results suggest that, although the number of publications authored by women has increased in the past two decades, women are still severely underrepresented in CS research. Our results also indicate that the main issue is not a smaller output of works by female authors, but that there are few women participating in CS research.

To help improve this scenario, different initiatives were created in Brazil to encourage women to join CS communities. The Brazilian Computer Society, for example, sponsors the Meninas Digitais\footnote{\url{https://meninas.sbc.org.br/}} program, which aims to promote technology to girls in elementary school and high school and encourage them to pursue a career in IT.

Some limitations of this research were related to the small sample of data from the analyzed conferences that were indexed in popular CS databases. The Brazilian Computer Society makes available most, if not all, of their conference publications on SBCOpenLib, but unfortunately SBCOpenLib does not have an integrated API to facilitate this type of systematic research. 

Another challenge presented was the use of gender inference services, which erase other gender identities by forcing a binary parameter of male or female. To advance more inclusive and accurate research on diversity and representation of minorities, one alternative would be to collect demographic information such as gender by asking for the author's self-identification and including it with the publication's metadata.

As future work, we intend to expand this research with a larger data set and conduct further statistical analysis, as well as compare the results for Brazilian conferences with other international conferences. Most importantly, we aim to use this research as a reference for future analysis of the CS field, evaluating how the current initiatives and efforts being made to encourage women to join CS might impact the gender balance in the future.

The discussion about female representation goes beyond the ethical responsibility of ensuring more equitable gender representation in the future. Diversity is key to advance CS research, as different backgrounds and life experiences present an advantage by giving individuals unique insights and approaches to problem-solving. We hope that this study prompts other works with a focus on analyzing gender disparities in Brazil, and encourages reflection by the community members about the cause of such problems and possible strategies to increase diversity in the Computer Science field moving forwards.

\bibliographystyle{unsrt}  
\bibliography{references}

\end{document}